\documentclass[openacc]{rstransa}

\usepackage{amssymb}
\usepackage{bm}

\def\mr#1{{\textcolor{black}{#1}}}

\begin{document}

\title{Puff turbulence in the limit of strong buoyancy}

\author{
Andrea Mazzino$^{1}$ and Marco Edoardo Rosti$^{2}$}

\address{$^{1}$Department of Civil, Chemical and Environmental Engineering (DICCA), University of Genova, Via Montallegro 1, 16145 Genova, Italy; INFN, Genova Section, Via Montallegro 1, 16145 Genova, Italy\\
$^{2}$Complex Fluids and Flows Unit, Okinawa Institute of Science and Technology Graduate University, 1919-1 Tancha, Onna-son, Okinawa 904-0495, Japan}

\subject{xxxxx, xxxxx, xxxx}


\corres{Andrea Mazzino\\
\email{andrea.mazzino@unige.it}\\
Marco E. Rosti\\
\email{marco.rosti@oist.jp}}

\begin{abstract}
We provide a numerical validation of a recently proposed phenomenological theory to characterize the space-time statistical properties of a turbulent puff, both in terms of bulk properties, such as the mean velocity, temperature and size, and scaling laws for velocity and temperature differences both in the viscous and in the inertial range of scales. In particular, apart from the more classical shear-dominated puff turbulence, our main focus is on the recently discovered new regime where turbulent fluctuations are dominated by buoyancy. The theory is based on an adiabaticity hypothesis which assumes that small-scale turbulent fluctuations rapidly relax to the slower large-scale dynamics, leading to a generalization of the classical Kolmogorov and Kolmogorov-Obukhov-Corrsin theories for a turbulent puff hosting a scalar field. We validate our theory by means of massive direct numerical simulations finding excellent agreement.
\end{abstract}


\begin{fmtext}

\end{fmtext}


\maketitle

\section{Introduction}
Turbulence in a fluid puff is a problem recently attracting a great deal of attention in relation to the global emergency caused by the COVID-19 pandemy. Its relevance stems from the key role of airborne virus transmission by viral particles released by an infected person via coughing, sneezing, speaking or simply breathing \cite{lincei}. Focusing on a cough for the sake of example, its typical duration is 200-500 $ms$, the typical mouth opening of male subjects is $\sim$ 4 $cm^2$, and the resulting Reynolds number is about $10^4$ \cite{Bo20,gupta2009flow} and even larger for a sneeze. The resulting flow field is thus fully turbulent. Very recently \cite{Ro20, Ro21, Lo21}, the key role of turbulence in dictating the fate of virus-containing droplets in violent human expulsions has been elucidated. More specifically, it turns out that turbulence plays a crucial role in determining droplets evaporation time, resulting in errors up to 100 $\%$ when the turbulent fluctuations are filtered or completely averaged out \cite{Ro20}. Because of the fact that the issue on the droplet evaporation time is crucial to establish whether viruses lingering on dry nuclei upon droplet evaporation retain their full potential of infection \cite{Ro21}, turbulence in a puff needs to be fully understood and characterized in a statistical sense.

As a matter of fact, a consistent statistical theory is currently not available for puff turbulence. The reason is probably due to the fact that the system at hand does not possess all desirable symmetries characterizing the classical ideal turbulence setting. Current knowledge of puff turbulence is confined to the pioneering work by Kovasznay et al. (1975) \cite{Ko75} only dealing with bulk quantities involved in the puff dynamics: the puff bulk translational velocity and the average puff radius.

A first attempt to characterize from a statistical point of view the small-scale structure of turbulent fluctuations of both velocity and temperature, but neglecting the two-phase nature of the problem, is reported in Ref.\ \cite{PRL} where an adiabatic generalization of the Kolmogorov-Obukhov picture of steady Navier-Stokes turbulence \cite{Ko41,Ob41} has been presented in two relevant cases: one in which the buoyancy of the puff is negligible and the opposite situation where buoyancy dominates the puff dynamics. It is shown that the phenomenological theory gives sense to the concept of inertial range, energy flux and scaling behaviors, in space and time, of proper statistical observables. A similar approach have been successfully used also in describing inertial range scaling laws in decaying homogeneous anisotropic turbulence \cite{Bi03}.

The proposed theory is still waiting for a verification (in terms of accurate DNS and/or experiments) in the second limit dominated by buoyancy. In Ref.\ \cite{PRL} only the case of negligible buoyancy and the intermediate regime where buoyancy is just starting to play a role have been analyzed. Here, we fill the gap by analyzing, in terms of state-of-the-art DNS, the buoyancy-dominated regime with the aim of supporting the theoretical prediction of Ref.\ \cite{PRL}.

The paper is organized as follow. In Sec.~\ref{sec:eq} we describe the governing equations governing a turbulent puff. In Sec.~\ref{sec:theory} we propose a phenomenological model to describe the spatial and temporal evolution of the turbulent puff, both in terms of bulk properties (mean velocity, temperature and size) and velocity differences in the inertial and viscous range of scales; in order to test the validity of the theory, we perform Direct Numerical Simulations, whose details are provided in Sec.~\ref{sec:dns}. The comparison of theory and simulations is reported in Sec.~\ref{sec:res}, and the main conclusions are summarized in Sec.~\ref{sec:end}.

\section{Problem setting and ruling equations} \label{sec:eq}
\begin{figure}
\centering
\includegraphics[width=0.65\textwidth]{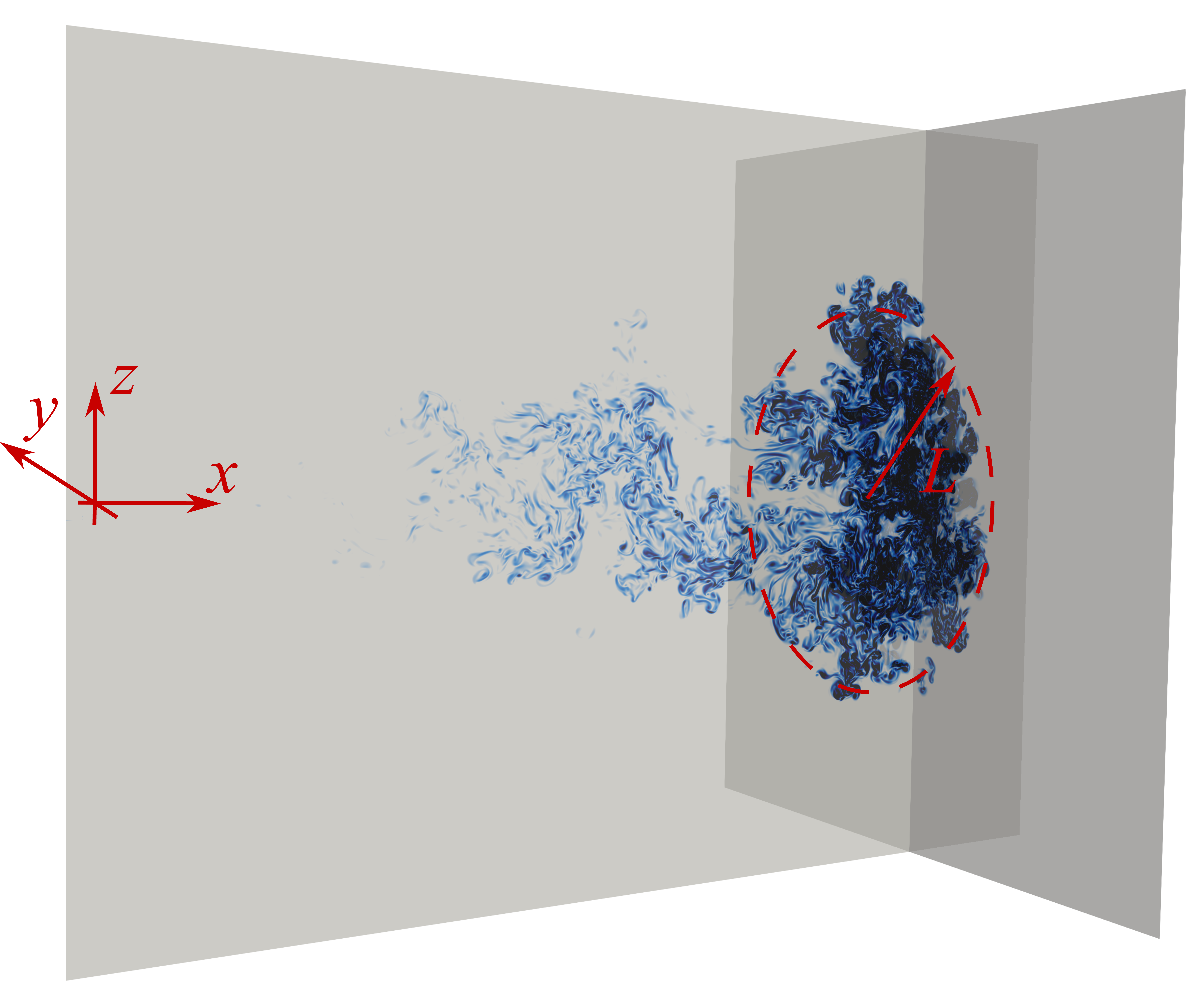}
\caption{Side view and cross section of the turbulent puff studied in the present work, together with the reference coordinate axis.}
\label{fig:sketch}
\end{figure}
We consider a fluid impulsively ejected in an undisturbed environment from a small circular opening of radius $\mathcal{R}$, as shown in Fig.~\ref{fig:sketch}. The fluid motion is governed by the incompressible Navier-Stokes equations for the velocity $\bm{u}$ and pressure $p$ fields, coupled to the advection-diffusion equation for the temperature field $\theta$, i.e.\ by the Oberbeck-Boussinesq (OB) equations \cite{Tr88}: 
\begin{equation}
  \partial_t \bm{u} + \bm{u}\cdot \bm{\partial} \bm{u}=-\frac{\bm{\partial}p}{\rho_a} + \nu \partial^2 \bm{u} -\beta \bm{g} (\theta-\theta_{a}),
  \label{eq:NS}  
\end{equation}
\begin{equation}
  \bm{\partial}\cdot \bm{u}=0,
  \label{eq:div0}  
\end{equation}
\begin{equation}
  \partial_t \theta + \bm{u}\cdot \bm{\partial} \theta = \kappa \partial^2 \theta.
  \label{eq:T}
\end{equation}
Here,  we assumed the fluid density $\rho $ to be linearly dependent on the temperature $\theta$: $\rho(\theta)=\rho_a[1-\beta (\theta-\theta_a)]$, $\theta_a$ and $\rho_a$ being the (constant) ambient temperature and density, respectively,  and $\beta$ the thermal expansion coefficient. In the previous equations, $\bm{g}=(0,0,-g)$ is the gravitational acceleration, $\kappa$ the thermal diffusion coefficient and $\nu$ ($\sim \kappa$) the kinematic fluid viscosity.  In the following, we denote $T$ as the difference between the fluid temperature and the ambient one: $T=\theta-\theta_a $. Note that, the momentum equation for the velocity field and the temperature advection-diffusion equation are fully coupled; indeed, the temperature field does not behave as a simple passive scalar field advected by the velocity field, but can alter the flow field itself by the back-reaction term $\beta \bm{g} (\theta-\theta_{a})$ in the momentum equation.

\section{Phenomenological theory for bulk and two-point puff statistics} \label{sec:theory}

In this section we provide a quick review of known results for both bulk puff properties (e.g. typical puff velocity and typical puff size) and two-point observables characterizing the mixing zone of the puff. The results we are going to review for the bulk properties in the limit of vanishing buoyancy are known since the seminal paper \cite{Ko75}. The remaining results have been presented only very recently by the present authors \cite{PRL} and are still waiting for a conclusive confirmation from numerical simulations and/or experiments. Providing such a confirmation in terms of state-of-the-art DNS is the main concern of the present paper.

Let us consider a fluid cloud impulsively ejected from a spatially-localized source (e.g. at the origin of a 3D reference system) in an otherwise undisturbed environment (e.g. air). The injection process is supposed to last until time  $t_0$, after which  the cloud freely evolves into the ambient under the combined action of gravity and shear. In the freely evolving regime, the fluid cloud is named puff, to distinguish it from the initial jet phase \cite{Gh10}. Let us denote by $T_0>0$ the difference between the cloud temperature and the ambient temperature at the end of the jet phase.

When the buoyancy effect is negligible, the temporal behavior of the typical size of the puff, denoted by $L(t)$, animated by a typical velocity $u_L\sim L/t$, follows from the momentum conservation law of the puff, $L^3 u_L=const$; \mr{note that, the typical translational velocity of the puff is determined in a Lagrangian frame of reference, following the point associated to the maximum velocity of the puff}. Namely: $L(t)\sim L_0 (t/t_0)^{1/4}$ and thus $u_L(t)\sim  u_0 (t/t_0)^{-3/4}$. Here, $L_0$ and $u_0=L_0/t_0$ are the typical cloud size and cloud velocity at the end of the jet phase. These scaling behaviors have been obtained in \cite{Ko75} from the large-scale Navier-Stokes equations. These latter indeed reduce to a diffusive equation with a time dependent eddy viscosity $\nu_T(t)\sim (t/t_0)^{-1/2}L_0^2/t_0 $.  The scaling laws of the puff bulk properties thus follow from simple power counting. Once the scaling behavior of $L(t)$ is known, one can easily predict the power-law decay for the large-scale temperature difference, $T_L$, between the puff and the cooler ambient. $T_L$ is indeed proportional to the reciprocal of the cloud volume $L^3$, from which: $T_L\sim T_0 (L/L_0)^{-3} = T_0 (t/t_0)^{-3/4}$.  Note that,  the theory discussed above is based on a time-dependent eddy-viscosity closure and predicts the scaling behavior for the (sole) bulk properties. As a results,  we found the same scaling laws for the typical radius of the puff and for the distance from the origin traveled by the puff.  It then follows that, apart prefactors, lengths in a puff obey the same scaling behaviors and the same applies also for the typical fluctuating velocities inside the puff and the mean velocity of the puff. They are indeed dimensionally related to the typical radius of the puff and to the distance it travels from the origin.

The eddy-viscosity description obtained in \cite{Ko75} makes it possible to generalize the above scaling behavior to the buoyancy-dominated case. Indeed, the order of magnitude of the buoyancy contribution to the puff evolution is $b=\beta g T_L$ where $\beta$ is the air thermal expansion coefficient and $g$  is the gravitational acceleration. For $T_L\sim  T_0(t/t_0)^{-3/4}$ one has $b\sim \beta g T_0(t/t_0)^{-3/4}$.  Because the eddy viscosity term goes as $t^{-7/4}$, soon or later buoyancy dominates (i.e. its decay is slower)  over the eddy viscosity. This happens provided that $t\gg t_b\equiv u_0 /(\beta g T_0)$. For $t\gg t_b$ the acceleration on the left-hand-side of the Navier--Stokes equations now balances the buoyancy term: $u_L/t\sim u_L^2/L \sim \beta g T_0 (L/L_0)^{-3}$ originating the following new scaling laws \cite{PRL}:

\begin{equation}
 L(t)\sim L_0  (t_0/t_b)^{1/4} (t/t_0)^{1/2}\qquad u_L  \sim u_0 (t_0/t_b)^{1/4} (t/t_0)^{-1/2}\qquad T_L\sim  T_0 (t_b/t_0)^{3/4} (t/t_0)^{-3/2}.
\label{newscal}
\end{equation}

Let us now move to analyze the two-point statistics and assume that the bulk quantities we have identified serve as large-scale properties of the flow, below which a cascade process may originate according to a generalized Kolmogorov-Obukhov picture. The idea is that, despite the fact that the problem at hand is not stationary, and thus very far from the classical playgroud of the Kolmogorov theory, small-scale turbulent fluctuations can rapidly relax to the slower large-scale dynamics. This is the essence of the adiabaticity hyphothesis leading to a generalization of the classical Kolmogorov theory for a turbulent puff. We refer to \cite{PRL} for the phenomenological theory when the buoyancy is negligible.  Here we focus our attention on the sole regime where buoyancy dominates and the bulk puff properties are ruled by the scaling laws (\ref{newscal}). The starting point is to assume the existence of an inertial range of scales characterized by a constant flux of energy given by $\epsilon(t)\sim u_L^3/L\sim (\delta_r u)^3/r $. Exploiting the scaling behaviors (\ref{newscal}) one immediately gets $\epsilon(t)\sim \epsilon_0 (t_0/t_b)^{1/2} (t/t_0)^{-2}$ and

\begin{equation}
  \delta_r  u (t)\sim \epsilon_0^{1/3}r^{1/3}
  \left ( \frac{t_0}{t_b}\right )^{1/6} \left ( \frac{t}{t_0}\right )^{-2/3},
    \label{u-buoy2}
\end{equation}
where $\epsilon_0=u_0^3/L_0$ is the energy flux a $t=t_0$.  It is worth observing that to get Eq.~(\ref{u-buoy2}) we have supposed that buoyancy is only important to dictate large-scale balance, being instead negligible within the inertial range.  A similar scenario has been found to hold in other convective systems as, e.g.,  Rayleigh-Taylor turbulence \cite{Ch03,Vl09,Bo09,Ce06,Bo17}.

The energy flux is supposed to persist up to the generalized Kolmogorov viscous scale, $\eta$: the scale at which the inertial self-advection matches the viscous term in the Navier--Stokes equations. Namely, $\delta_{\eta} u\  \eta \sim \nu$. Evaluating $\delta_{r} u$ at $r=\eta$ in Eq.~(\ref{u-buoy2}) the expressions for $\eta$ follows:
\begin{equation}
 \eta (t)\sim \nu^{3/4}\epsilon_0^{-1/4} \left ( \frac{t_0}{t_b}\right )^{-1/8} \left ( \frac{t}{t_0}\right )^{1/2}.
    \label{eta}
\end{equation}
Below $\eta$ velocity fluctuations are smooth,
i.e.\ $\delta_r u \sim (r/\eta)\delta_{\eta} u$, and the resulting scaling behavior for $r\ll \eta$ is:
\begin{equation}
  \delta_r u (t)\sim r \left (\frac{\epsilon_0}{\nu}\right )^{1/2} \left ( \frac{t_0}{t_b}\right )^{1/4} \left ( \frac{t}{t_0}\right )^{-1} .
  \label{u-viscous-buoy}
\end{equation}

A similar way of reasoning for the two-point temperature fluctuations, $\delta_r T$, leads to the adiabatic generalization of Obukhov-Corrsin theory (OC51) of passive scalar advection \cite{Ob49,Co51}:
\begin{equation}
  \delta_r T (t)\sim \varepsilon_0^{1/2}\epsilon_0^{-1/6} r^{1/3} \left ( \frac{t_b}{t_0}\right )^{5/6} \left ( \frac{t}{t_0}\right )^{-5/3},
\label{T-buoy2}
  \end{equation}
where  $\varepsilon_0=u_0 T_0^2/L_0$ is the flux of scalar variance at $t=t_0 $.
The above scaling law is accompanied by the scaling relation within the temperature dissipative range (i.e.\ for scales $r\ll r_d$, $r_d$ being the dissipative scale coinciding with $\eta$ because here $\nu\sim\kappa$):
\begin{equation}
  \delta_r T (t) \sim r \left ( \frac{\varepsilon_0}{\nu}\right )^{1/2} \left ( \frac{t_0}{t_b}\right )^{-3/4} \left ( \frac{t}{t_0}\right )^{-2}.
  \label{T-diff-buoy}
\end{equation}

The scaling behaviors (\ref{u-buoy2}) and (\ref{T-buoy2}) fix, via simple power-counting, the mean field predictions for the equal-time velocity and temperature inertial range structure functions. To account for intermittency corrections of both velocity \cite{Fr95} and temperature fluctuations \cite{Ve97,Fr98,Sh00,Fa01} a further assumption is needed. Accordingly, in \cite{PRL} it has been postulated that a turbulent puff possesses the same spatial scaling exponents as those of the stationary, homogeneous and isotropic turbulent system hosting a passively behaving scalar. This was expected to hold coherently with the adiabaticity hypothesis. The mean-field predictions for the  structure functions of order $p$ are thus corrected by a multiplicative factor $[r/L(t)]^{-\sigma_{p}}$, for the velocity, and $[r/L(t)]^{-\xi_{p}}$ for the temperature. Note that because of the temporal dependence encoded in $L(t)$ the intermittency corrections change both the spatial and the temporal structure-function scaling laws. The values of $\sigma_{p}$ and $\xi_p$ used are those of the stationary, homogeneous and isotropic turbulence advecting a scalar field \cite{PRL} taken from \mr{state-of-the art numerical simulations of homogeneous isotropic turbulence} \cite{Wa07}; in particular, we use \mr{$\sigma_4=0.06$, $\sigma_6=0.27$, $\xi_4=0.24$ and $\xi_6=0.37$}. The resulting model for the structure functions of the (longitudinal) velocity and the temperature difference is thus:
\begin{equation}
 S^{\parallel}_p(r)=A\, \epsilon_0^{p/3}r^{p/3-\sigma_p}
  \left ( \frac{t_0}{t_b}\right )^{p/6+\sigma_p/4} \left ( \frac{t}{t_0}\right )^{-2 p /3+\sigma_p /2} L_0^{\sigma_p},
  \label{sfu:buoy.final}
\end{equation}
and
\begin{equation}
  S_{2p}(r)= B\,\varepsilon_0^{p}\epsilon_0^{-p/3} r^{2 p/3-\xi_{2p}} \left ( \frac{t_b}{t_0}\right )^{5 p/3-\xi_{2p}/4} \left ( \frac{t}{t_0}\right )^{-10 p/3+\xi_{2 p}/2} L_0^{\xi_{2p}},
  \label{sfT:buoy.final}
\end{equation}
where $A$ and $B$ are unknown, non-universal (i.e.~dependent on all details of the system) prefactors.The predictions for the structure functions within the viscous/dissipative range of scales follow from (\ref{u-viscous-buoy}) and (\ref{T-diff-buoy}) by simple power counting without any intermittency correction.

In the above discussion, we assumed that the coupling between velocity and temperature is negligible everywhere in the inertial range of scales, except at the largest scales where buoyancy fixes the balance between inertia and buoyancy (i.e.  it forces the velocity fluctuations).  Our assumption led to a Kolmogorov scenario which is consistent with the hypothesis that temperature is passive within the inertial range of scales. If one assumes that velocity and temperature are scale-by-scale balanced within the inertial range of scales, a new scaling behavior would emerge (scaling a la Bolgiano), which is however not observed in our simulations.  In conclusion, the fact that temperature behaves passively in the inertial range of scales is a work hypothesis which is successively verified via a consistency check.

\section{Direct numerical simulations} \label{sec:dns}
In order to verify the above phenomenological theory, we rely on fully resolved direct numerical simulations. The equations of motion (\ref{eq:NS})--(\ref{eq:T}) are solved numerically within a cubic domain box of size $\mathcal{L}=84 \mathcal{R}$, $\mathcal{R}$ being the radius of the circular opening from which air is injected according to the time-varying profile representative of human cough proposed by Gupta et al.\ \cite{gupta2009flow}, with a peak Reynolds number of $Re=u_{max} 2 \mathcal{R}/\nu \approx 20000$.  The flow is assumed to be periodic at the four sides and we prescribe a convective outlet boundary condition at the outflow boundary.  We use the flow solver \textit{Fujin}, an in-house code, extensively validated in a variety of problems~\cite{rosti_brandt_2017a, rosti2019flowing, rosti_ge_jain_dodd_brandt_2019, Al19, devita2019, devita2020,rosti2020increase, olivieri2020dispersed}, based on the (second-order) finite-difference method for the spatial discretization and the (second-order) Adams-Bashfort scheme for the temporal discretization. See also: \texttt{https://groups.oist.jp/cffu/code}.

In the performed simulations, the domain is discretized with $1260$ grid points per side with a uniform spacing in all directions, resulting in a total number of $2$ billion grid points.  We verified the convergence of the results by comparing them with those obtained with different grid resolutions. The structure function presented below from the results of the simulations are obtained as follows: first we identify the distance from the emission point where the maximum velocity fluctuation is present. In that plane, and in $80$ neighboring ones ($40$ preceding and $40$ following it), we compute the velocity difference at the desired power and average the results. Note that, only the points inside the puff are considered; these are selected by choosing those in which the temperature is $1\%$ greater than the ambient one.

\section{Results} \label{sec:res}
\begin{figure}
\centering
\includegraphics[width=0.45\textwidth]{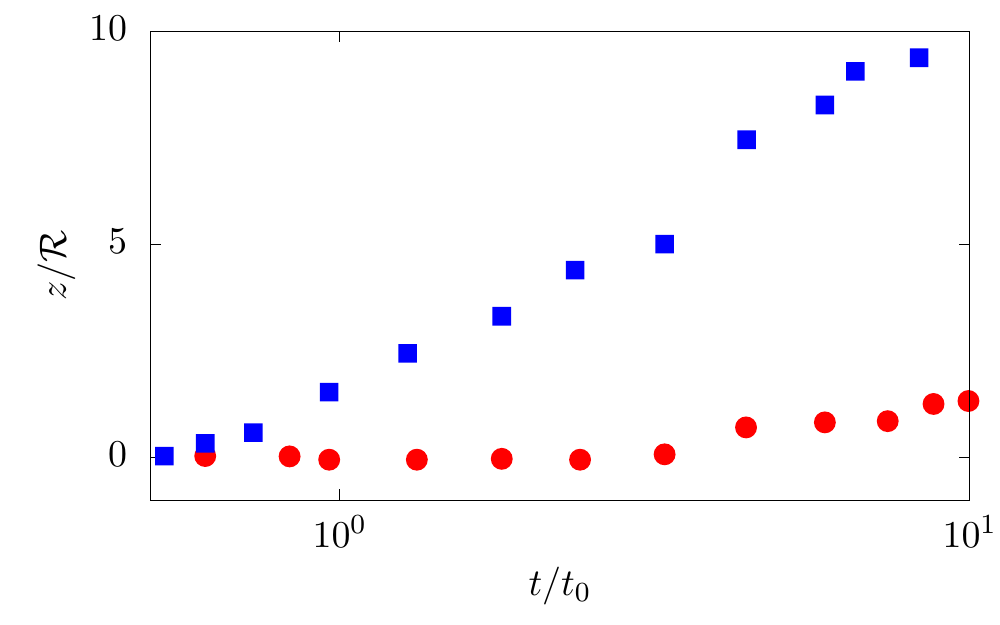}
\caption{Time history of the vertical displacement of the puff center of mass. The red circles and blue squares represent the data from the case with low and high buoyancy.}
\label{fig:cg}
\end{figure}
We start our analysis by showing in Fig.~\ref{fig:cg} the time evolution of the center of mass vertical position for two reference configurations: one where buoyancy is dominant (blue symbols) with $b \approx 0.5$ ($t_b \approx 2 t_0$) and one where its effect is secondary (red symbols) with $b \approx 0.05$ ($t_b \approx 14 t_0$). As shown in the figure, while in the  former the puff starts raising due to buoyancy soon after the ejection phase ends at $t=t_b \approx t_0$, the latter instead initially maintains an almost straight trajectory, and only at later times when $t \approx t_b \gg t_0$ starts rising. We can thus safely consider the two cases as good representative of the buoyancy-dominated and shear-dominated regimes.

\begin{figure}
\centering
\includegraphics[width=0.45\textwidth]{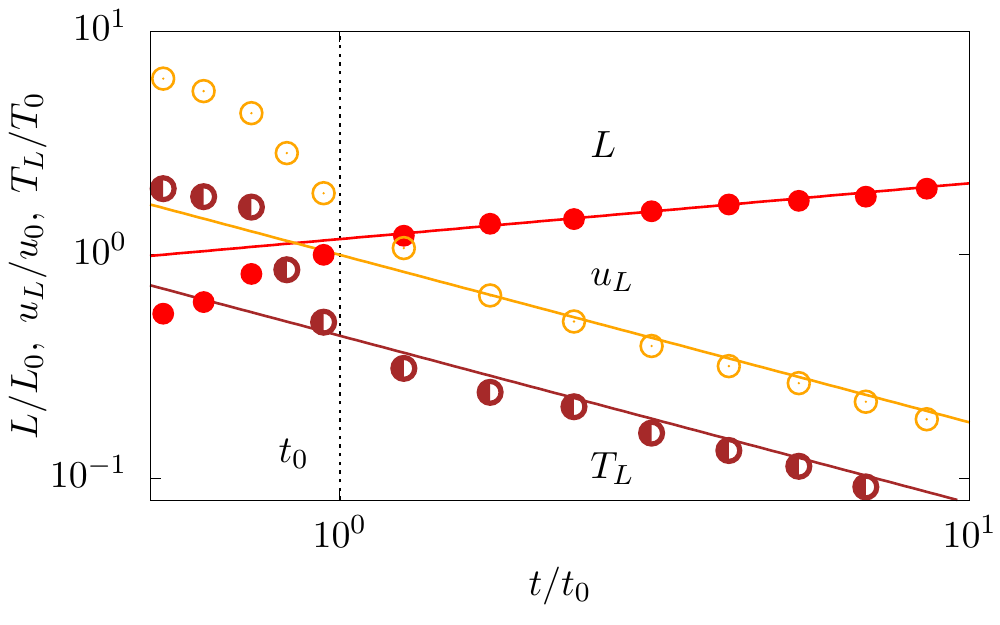}
\includegraphics[width=0.45\textwidth]{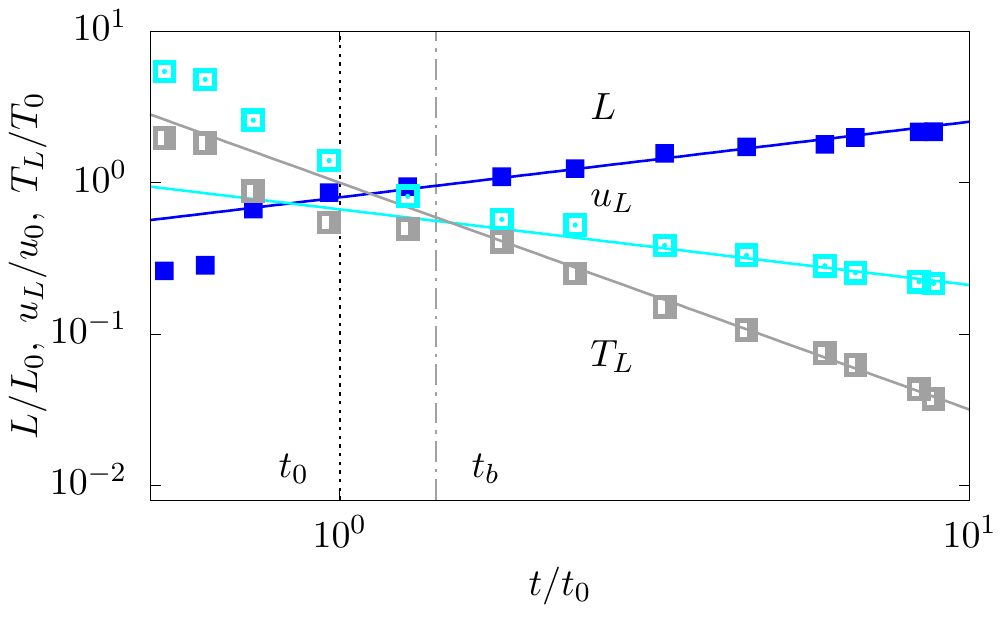}
\caption{Scaling laws for the bulk properties of the puff: $L$ (filled symbols), $u_L$ (empty symbols) and $T_L$ (half-filled symbols) (divided by a factor two for graphical reasons) for (left) shear-induced fluctuations and (right) buoyancy-driven fluctuations. The data reported in the left panel are taken from \cite{PRL}. In the figures, the solid lines represent the proposed scaling laws, while the symbols the results of our simulations.}
\label{fig:bulk}
\end{figure}
For the two configuration above, we start reporting in Fig.~\ref{fig:bulk} the bulk property of the puff evolution, namely its bulk size $L$, the streamwise velocity $u_L$ and the temperature difference $T_L$.  In both configurations, after the initial transient phase of the ejection which is dependent on the emission profile, the bulk properties of the puff attain power law behaviors. The puff size $L$ grows in size, and accordingly it slows down and adapt to the ambient temperature. In the figures we report the theoretical predictions with solid lines and the simulations data with symbols. As can be appreciated, the agreement between the two is very good. Note that, the theory is not expected to hold in the early stage of the evolution, but only at later stages after the information of the initial condition is lost. In particular, from the figure we note that the scaling laws for the shear-dominated regime are valid for $t \gtrsim t_0$, while those for the buoyancy-dominated regime for $t \gtrsim t_b$.  Furthermore, to derive the scalings for the bulk properties we assumed momentum conservation.; in order to verify the validity of this assumption, we can define a viscous time as $t_\nu \approx L^2/\nu$ from which we obtain in our case $t_\nu / t_b \approx 8000 \gg 1$. Thus, we conclude that up to $t_b$ the viscous contribution is negligible at large scales and the momentum is conserved with good accuracy.  Note that, after $t_b$, we do not make any assumption on momentum conservation.

\begin{figure}
\centering
\includegraphics[width=0.45\textwidth]{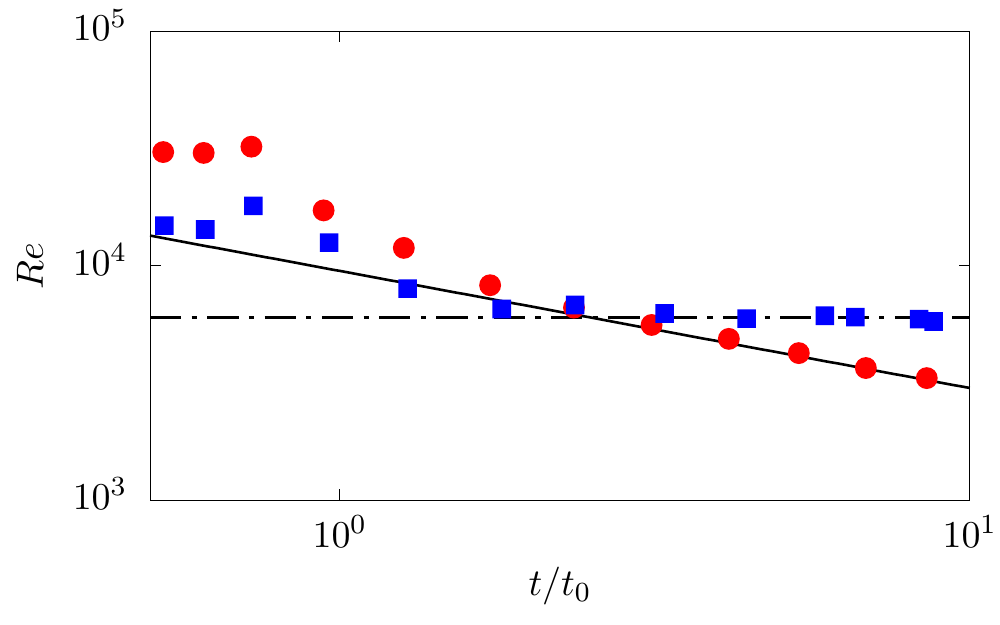}
\includegraphics[width=0.45\textwidth]{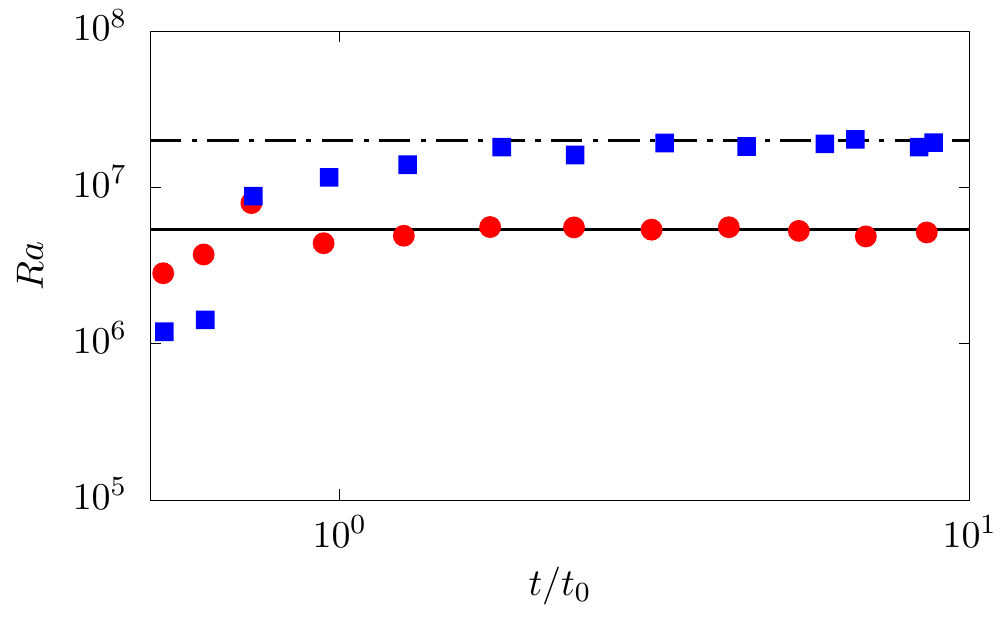}
\caption{Scaling laws for the non-dimensional numbers of the puff: (left) $Re$ and (right) $Ra$ for shear-induced fluctuations (red circles) and buoyancy-driven fluctuations (blue squares). In the figures, the solid lines represent the proposed scaling laws, while the symbols the results of our simulations.}
\label{fig:non-dim}
\end{figure}
A complete description of the bulk preperties of the puff can be obtained by considering the time evolution of two non-dimensional numbers: the Reynolds number $Re=u_L L/ \nu$ and the Rayleigh number $Ra= \beta g T_L L^3/ \left( \nu \kappa \right)$.  The time evolution of these quantities is shown in Fig.~\ref{fig:non-dim}, together with their theoretical predictions. We observe that both theory and numeric confirm that in the shear-dominated regime the Reynolds number $Re$ decays with time,  while the Rayleigh number remains constant over time. On the other hand, in the buoyancy dominated regime both are constant and independent of time.  These results further corroborates the validity of the phenomenological theory for the bulk properties of the flow. Based on these, we can now proceed to verify the predictions for the two-point statistics of the turbulent puff.

\begin{figure}
\centering
\includegraphics[width=0.45\textwidth]{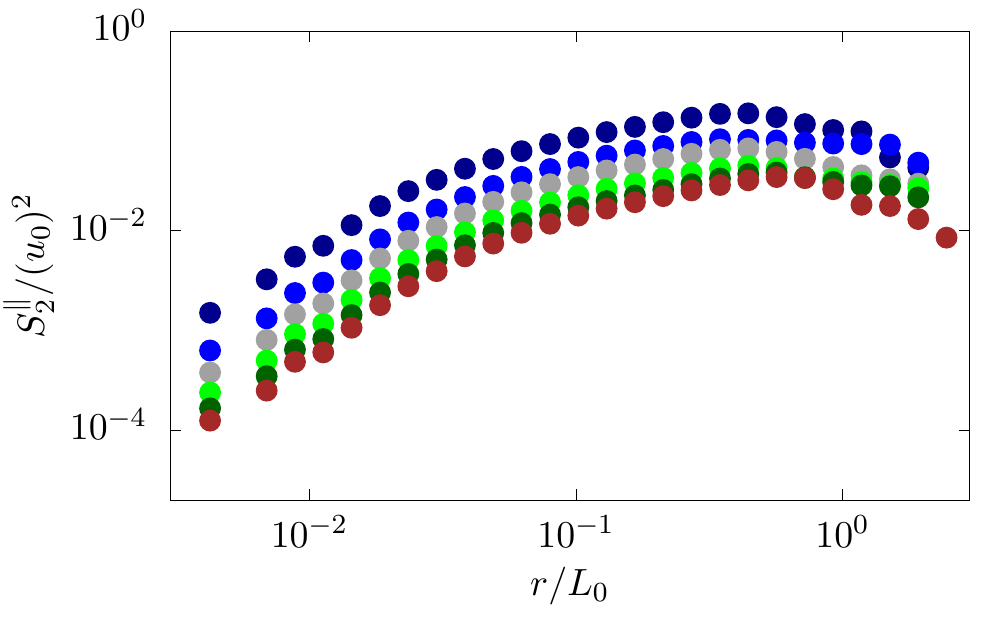}
\includegraphics[width=0.45\textwidth]{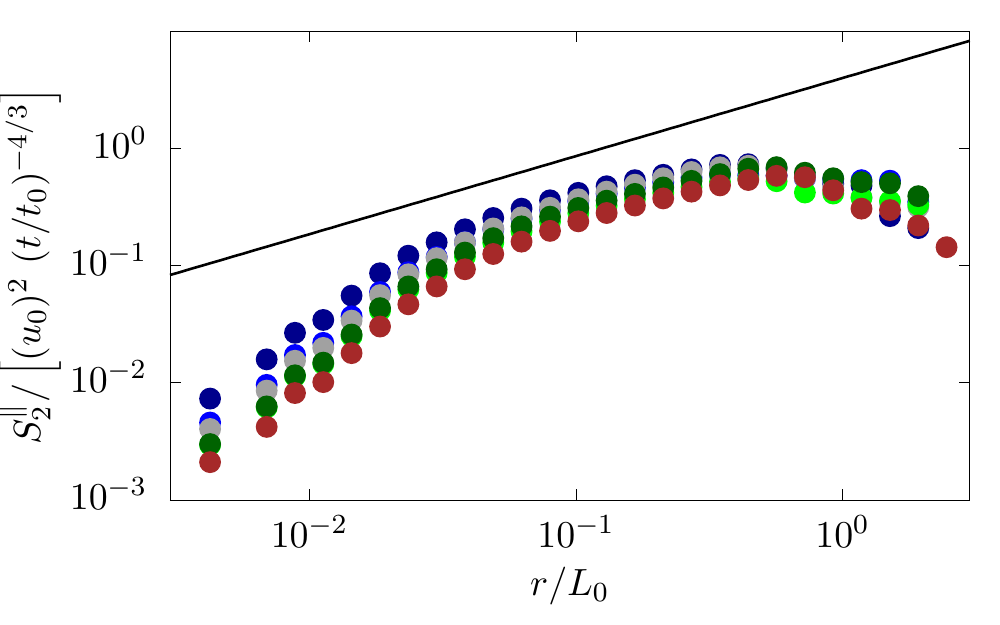} \\
\includegraphics[width=0.45\textwidth]{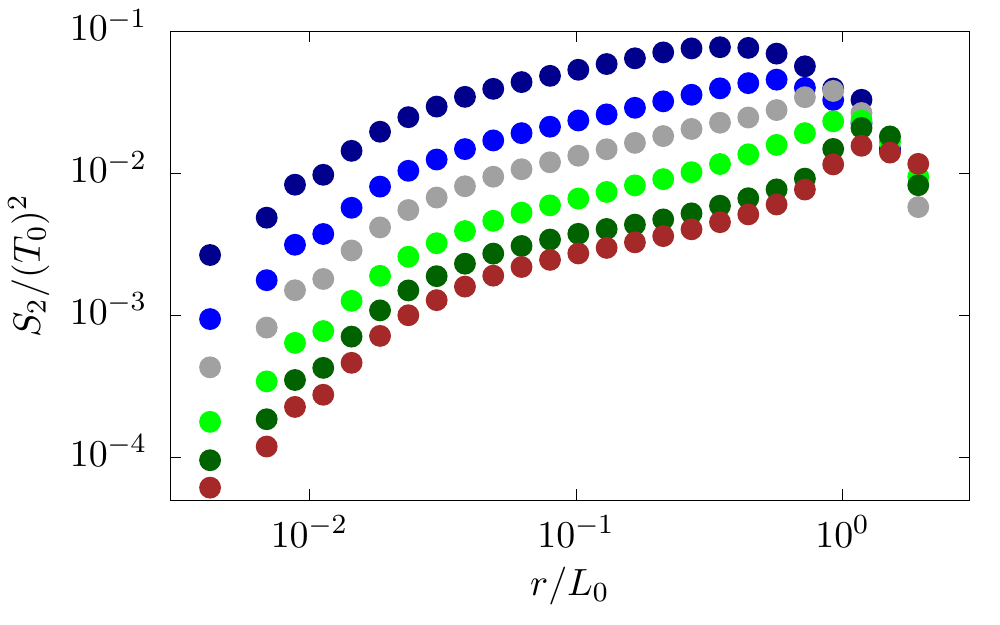}
\includegraphics[width=0.45\textwidth]{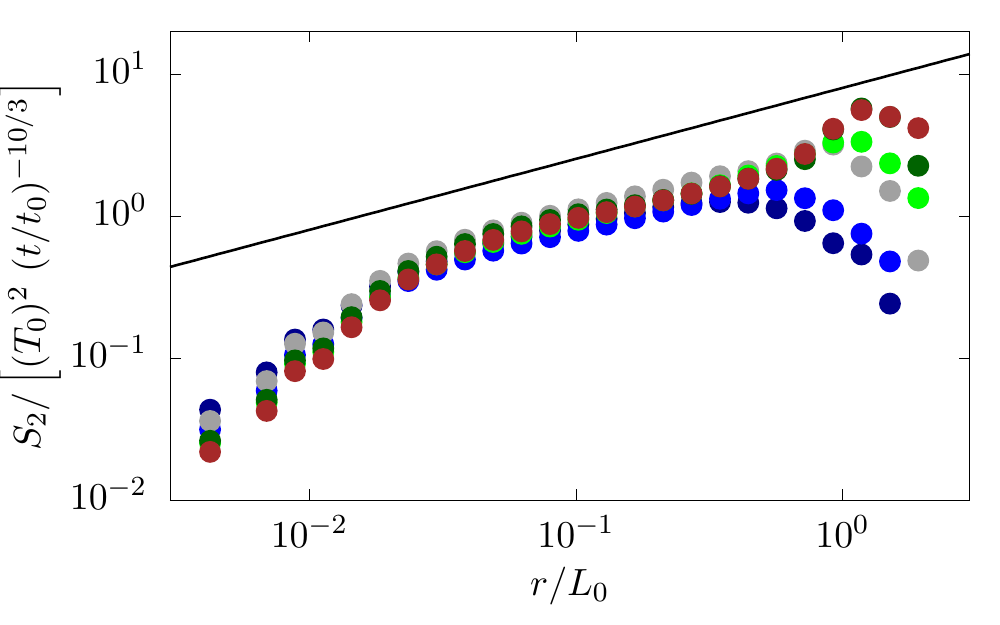}
\caption{Second order structure function for the (top) velocity and (bottom) temperature difference: (left) raw data and (right) data scaled by the predicted inertial scaling. In the figures, the solid lines represent the proposed scaling laws, while the symbols the results of our simulations. The data refers to the time interval $3 \lesssim t/t_0 \lesssim  10$, \mr{with colors indicating different time instants separated by an interval of $t_0$ in the following order: dark-blue, blue, gray, light-green, green and brown}.}
\label{fig:S2}
\end{figure}
Fig.~\ref{fig:S2} shows the second order structure function for the velocity and temperature differences for the time instants marked in Fig.~\ref{fig:bulk} after the end of the jet phase, i.e.~$t>t_0$. As can be seen by the left plots, the magnitude of the structure function decreases as time passes, providing a scattering of the curves. Also, consistently with the bulk predictions for this regime from Eq.~(\ref{newscal}), the variation in the temperature data is larger than what observed for the velocity structure function. The time variation can be successfully compensated by scaling the data with the theoretical predictions of Eqs.~(\ref{sfu:buoy.final}) and (\ref{sfT:buoy.final}), as reported in the right plots. When doing so, all the data reasonably collapse onto a single curve in the inertial range of scales, while remaining scattered in the viscous one. An opposite effect could be obtained by scaling the data with the theoretical prediction for the viscous range of scales (not shown).

\begin{figure}
\centering
\includegraphics[width=0.45\textwidth]{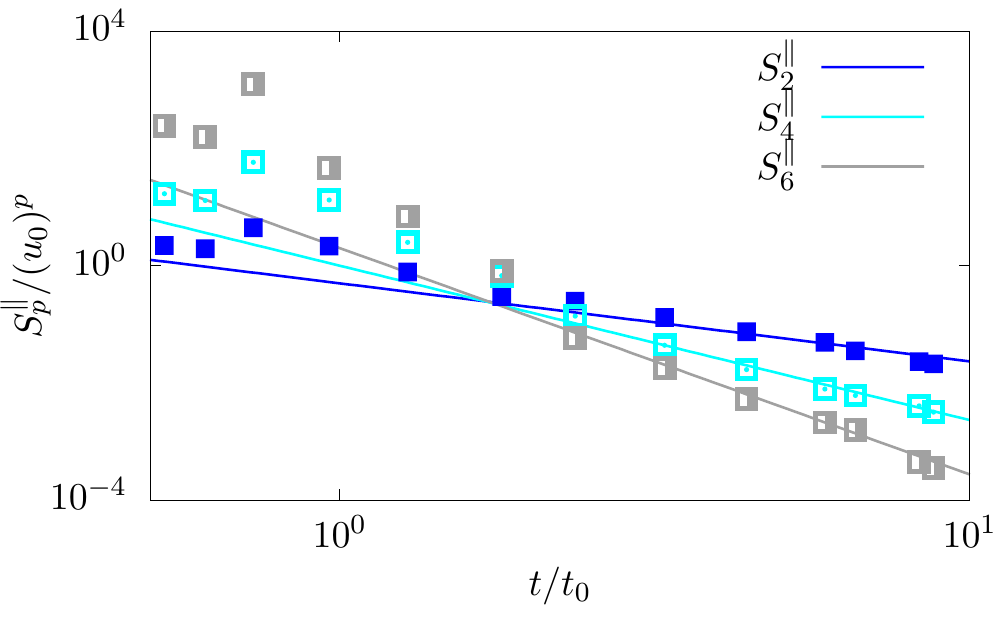}
\includegraphics[width=0.45\textwidth]{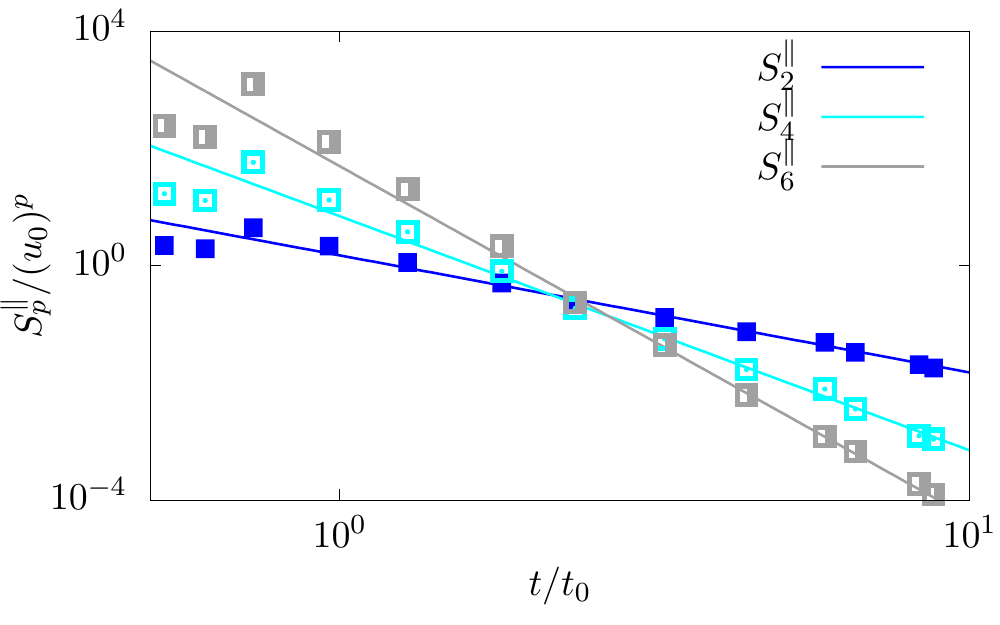}
\includegraphics[width=0.45\textwidth]{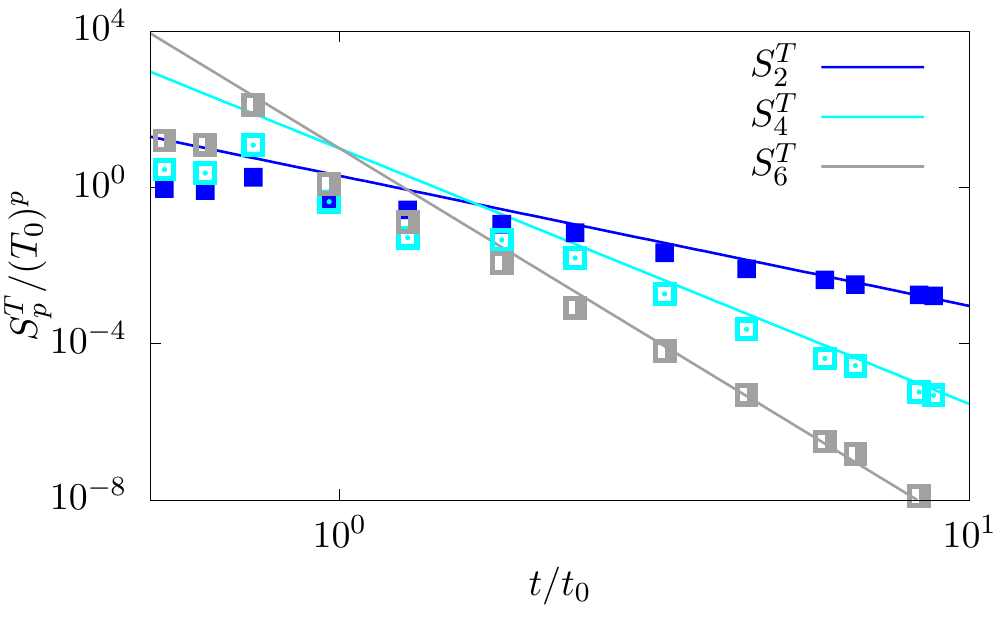}
\includegraphics[width=0.45\textwidth]{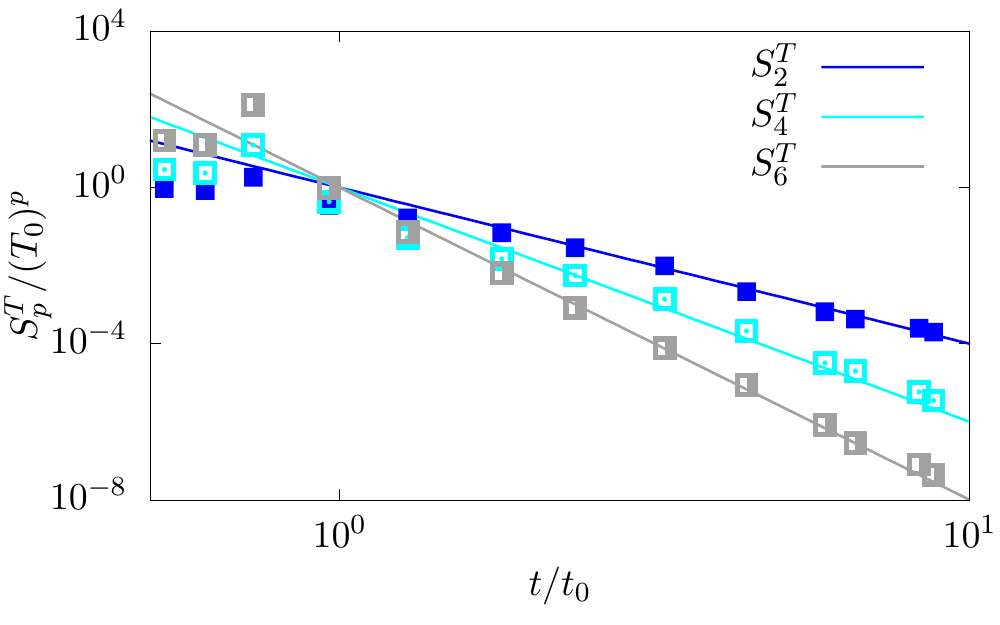}
\caption{Time histories of (top) $S_2^\parallel (r)$ (filled symbols), $S_4^\parallel (r)$ (empty symbols) and $S_6^\parallel (r)$ (half-filled symbols) and of (bottom) $S_2 (r)$ (filled symbols), $S_4 (r)$ (empty symbols) and $S_6 (r)$ (half-filled symbols) for two separations taken in the inertial (left) and viscous (right) range of scales, for the case with buoyancy-driven fluctuations. In the figures, the solid lines represent the proposed scaling laws, while the symbols the results of our simulations.}
\label{fig:time-2}
\end{figure}
As an additional step towards the verification of our scaling predictions for the structure functions in the viscous and inertial range of scales, we choose a separation $r/L_0$ in each of them ($r/L_0 \approx 0.007$ for the viscous range and $r/L_0 \approx 0.18$ for the inertial range) and report the time history of the second, fourth-order and sixth-order structure functions for the velocity and temperature difference, i.e.\ $S_2^\parallel (r)$, $S_4^\parallel (r)$, $S_6^\parallel (r)$ and $S_2 (r)$, $S_4 (r)$, $S_6 (r)$, respectively. These are shown in Fig.~\ref{fig:time-2} for the case with buoyancy-driven fluctuations. The continuous lines in the plots are the corresponding scaling laws from our theoretical predictions and we can observe in all cases an excellent agreement between the theoretical prediction and simulation data.  As stated above, for this flow the intermittency correction affect not only the spatial scalings but also the temporal ones reported here, thus the good agreement of our results is also suggesting that our predictions are correctly modeling the intermittency corrections.

\begin{figure}
\centering
\includegraphics[width=0.45\textwidth]{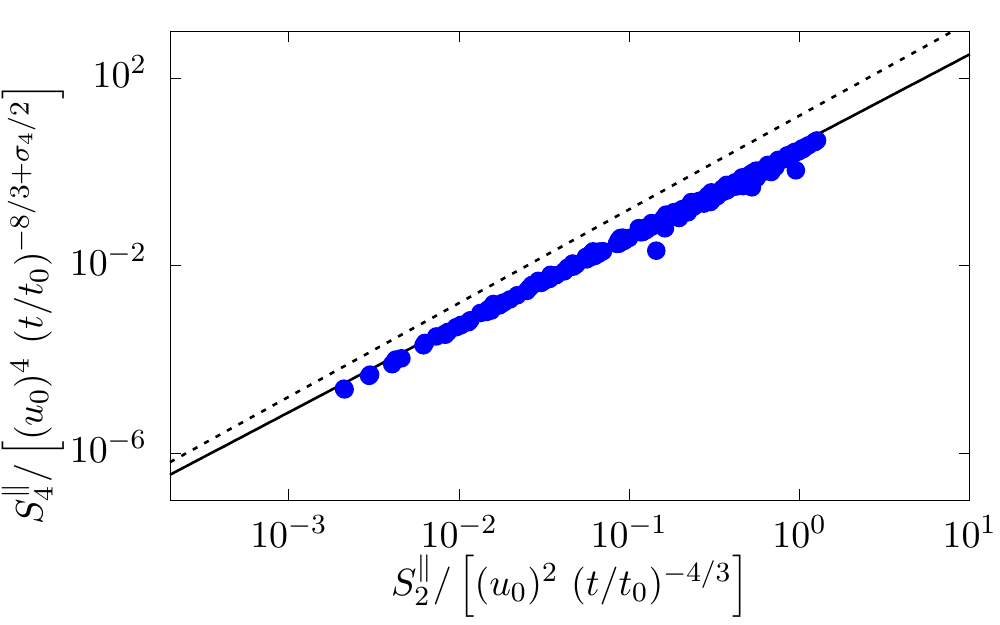}
\includegraphics[width=0.45\textwidth]{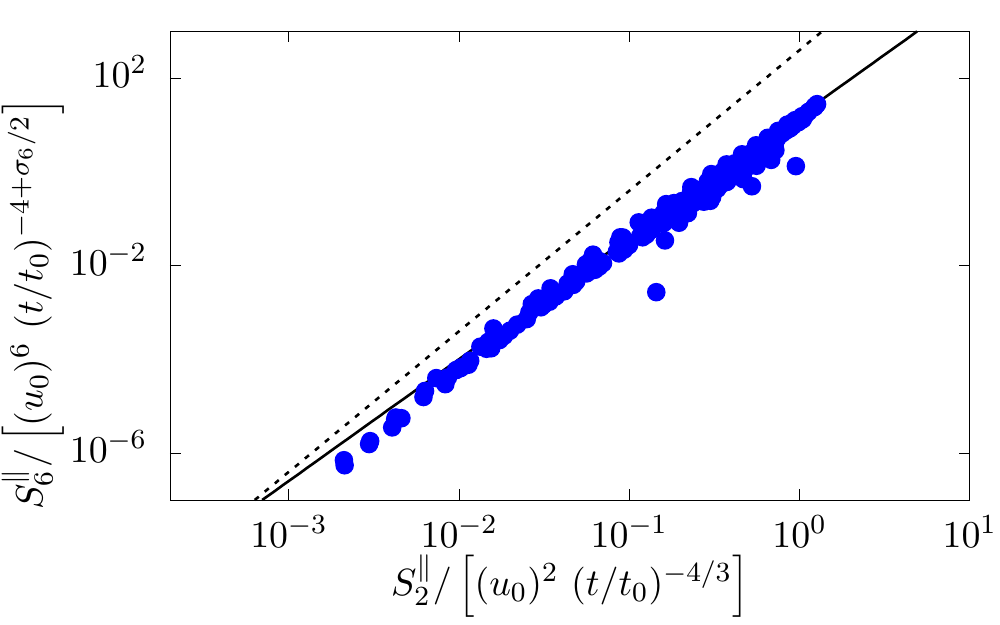}
\includegraphics[width=0.45\textwidth]{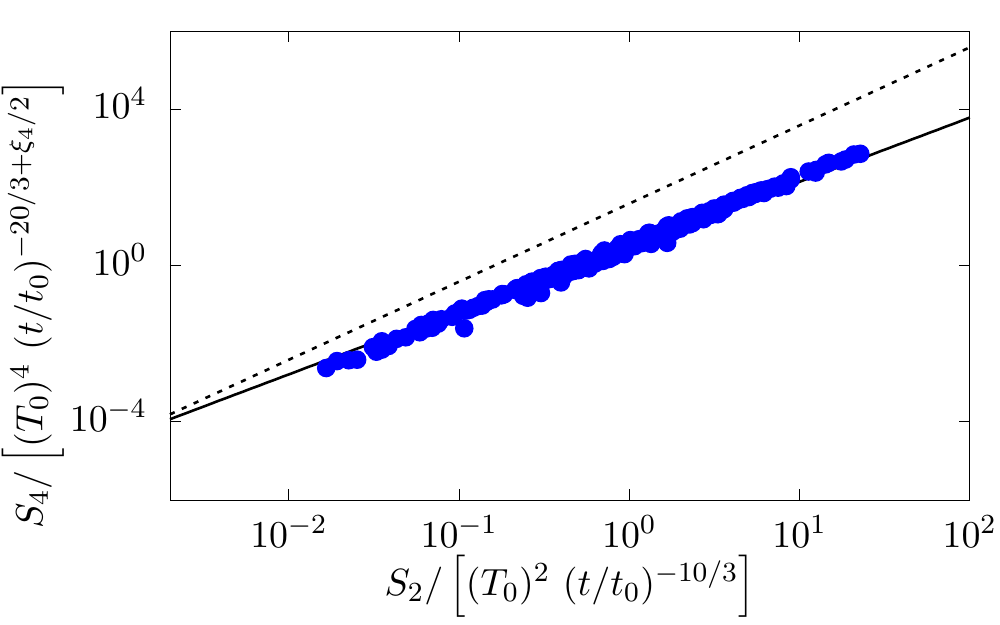}
\includegraphics[width=0.45\textwidth]{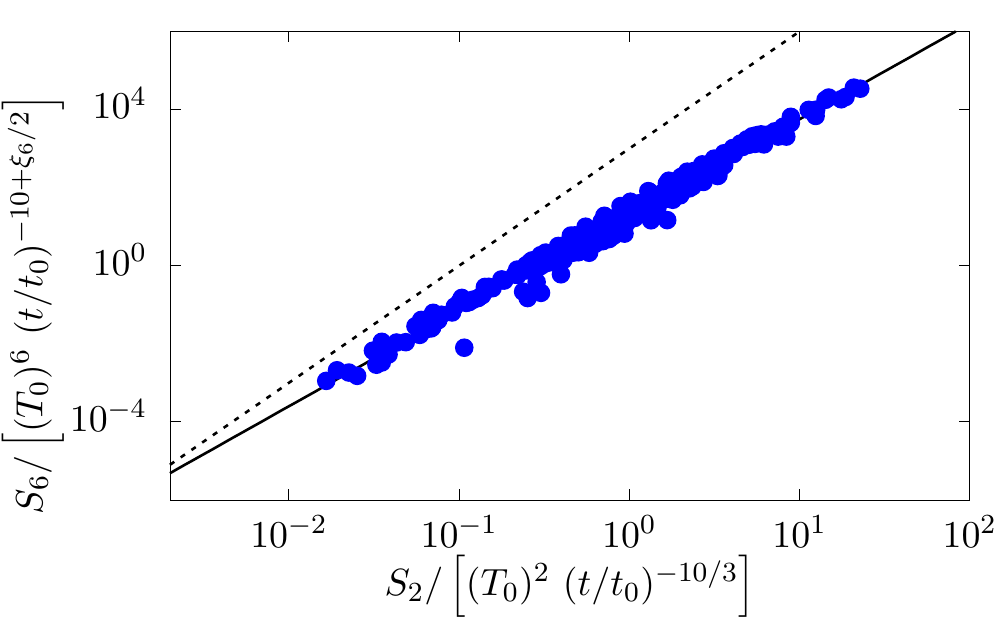}
\caption{(top) (left) $S_4^\parallel (r)$ and (right) $S_6^\parallel (r)$ as a function of $S_2^\parallel (r)$ and (bottom) (left) $S_4 (r)$ and (right) $S_6 (r)$ as a function of $S_2 (r)$.  All the structure functions are divided by their expected temporal scalings. The symbols represent the data from the case with high buoyancy, while the solid and dotted lines represent the ratio of the expected spatial scaling laws with and without the intermittency corrections.}
\label{fig:ess}
\end{figure}
Finally, we verify the theoretical predictions for both the temporal and spatial scalings altogether by showing in Fig.~\ref{fig:ess} the data in the form of Extended Self-Similarity \cite{Be93, Ch10}. Structure functions of any order in turbulence are expected to behave as power laws; Benzi and co-workers \cite{Be93} noted that even when the Reynolds number is not really large, any two structure functions show a mutual power law behavior, i.e.,~when plotted one against the other, a straight line appears with an exponent equal to the ratio between the two exponents., see e.g.~\cite{Ca95, Be96, Me96, Le99}  Here, we show the fourth and sixth structure functions for velocity and temperature  as a function of the second order one in the buoyancy dominated regime for several times following the end of the ejection (i.e.~$t>t_0$), the same of Fig.~\ref{fig:S2}. First, the good collapse of different time instants further corroborate the validity of our temporal scalings; next, the spatial one are verified by comparing the slope of the curve with the continuous lines with slopes deduced from our predictions, showing again a very good agreement. Note also that, the dashed lines are the resulting spatial scalings that do not account for the intermittency corrections, thus representing a pure mean field prediction; the clear departure of these lines from our data indicates both the importance of including intermittency corrections in the model, and the validity of our approach to do so \cite{Fr78}. 

\section{Conclusions} \label{sec:end}
A phenomenological theory for a turbulent puff has been recently presented in \cite{PRL}. The main focus of the theory is the two-point statistics of both velocity and temperature. This latter field is a scalar (turbulent) field carried by the puff turbulent velocity fluctuations. The theory, based on an adiabatic generalization of the classical Kolmogorov-Obukhov theory, identifies two distinct dynamical regimes. In the first, turbulence is mechanically driven and buoyancy only has a subleading role; the situation in reversed in the second regime where buoyancy drives turbulent  fluctuations. Associated to these different dynamical regime are different scaling laws, both in space and in time, for the two-point velocity/temperature structure functions.

While the numerical verification has been provided in \cite{PRL} for the mechanically-driven puff turbulence, the verification of the second regime was still lacking.  Here, we have provided such validation in terms of high-resolution DNS. As a result of our simulations, we first confirmed the existence of an inertial range of scales at the end of which a viscous (dissipative) region takes place. A similar conclusion has been drawn for the temperature field. The resulting phenomenology appears very similar to the classical Kolmogorov-Obukhov (and Kolmogorov-Obukhov-Corrsin for temperature fluctuations) scenario with turbulence fluctuations now relaxing almost instantaneously to the time-varying flow large-scale properties, with buoyancy effects only confined to the large-scale of motion.  This observation (which was a working-hypothesis in \cite{PRL}) led ourselves in \cite{PRL} to postulate the emergence of intermittency corrections in the present system as those known (numerically and experimentally) in the classical  homogeneous, isotropic and stationary turbulence. The validity of such model has been verified here with results clearly confirming the guess. We have also verified the space-time scaling behaviors of both velocity and temperature fluctuations resulting from the quasi-adiabatic scenario finding good agreement.

Our theory is limited to the case with $Pr \approx 1$ and this choice is motivated by the fact that air thermal diffusivity and viscosity are very close to each other.  From a theoretical point of view, the regime we have analyzed is also the most rich: indeed,  in this case an inertial range of scales for both velocity and temperature fluctuations exists, while this is not the case when $Pr \gg 1$ or $Pr \ll 1$.  In the former case, the scalar evolves in a smooth turbulent background without intermittency, the so-called Batchelor regime, for which the scaling behavior are reported in this work; in the latter case, despite an inertial range of scales develops for the velocity fluctuations, this is not for the scalar which exhibits a trivial bare diffusive regime.

As an interesting issue left for future research we mention the understanding of the buoyancy-dominated regime in two dimensions. This regime seems to be interesting because it might be characterized by a different dynamical regime with buoyancy now balancing, scale by scale, the nonlinear terms in the Navier--Stokes equations. A similar situation is known to happen in Rayleigh-Taylor turbulence and it causes the emergence of scaling laws \`a la Bolgiano (instead of \`a la Kolmogorov) \cite{Ma06}. Understanding whether or not this is the case also in puff turbulence is an interesting issue left for future.

The present results can help in future modelling efforts to predict the distance reached by viral load in human cough events. Indeed, it was found that droplet evaporation is mainly controlled by the combined effect of turbulence and droplet inertia \cite{Ro21}, and thus, the knowledge of the spatio-temporal \mr{evolution of turbulent fluctuations presented in this work can be used to obtain more realistic coarse-grained model of the original Navier-Stokes problem including the evaporation processes of small liquid droplets carried by the flow as, e.g., those involved in violent human expulsions} \cite{Ro20}. Another issue left for the future is the application and extension of the present theory to a multiphase problem.

\enlargethispage{20pt}


\dataccess{All data supporting the plots shown in the manuscript are available from the authors upon reasonable request.}

\aucontribute{A.M. and M.E.R. conceived the original idea and planned the research. M.E.R. developed the code and performed the numerical simulations. A.M. developed the theory. All authors analyzed data, outlined the manuscript content and wrote the manuscript.}

\competing{The authors declare that they have no competing interests.}

\funding{M.E.R. acknowledges the computational time provided by HPCI on the Oakbridge-CX cluster in the Information Technology Center, The University of Tokyo, under the grant hp200157 of the ``HPCI Urgent Call for Fighting against COVID-19" and the computer time provided by the Scientific Computing section of Research Support Division at OIST. A.M. thanks the financial support from the Compagnia di San Paolo, project MINIERA no. I34I20000380007, and from the Italian Ministry of University and Research (MUR),  project n.FISR2020IP-00290.}





\begin{thebibliography}{9}

\bibitem{lincei} G. Seminara, B. Carli, G. Forni, S. Fuzzi, A. Mazzino, and A. Rinaldo, Biological fluid dynamics of airborne COVID-19 infection. Rend. Fis. Acc. Lincei 31, 505 (2020).

\bibitem{Bo20} L. Bourouiba, Turbulent gas clouds and respiratory pathogen emissions: potential implications for reducing transmission of COVID-19, Jama 323, 1837–1838 (2020).

\bibitem{gupta2009flow} J.K. Gupta, C.H. Lin, Q. Chen, Flow dynamics and characterization of a cough. Indoor Air, 19(6):517–525 (2009).

\bibitem{Ro20} M. E. Rosti, M. Cavaiola, S. Olivieri, A. Seminara, and A. Mazzino. Turbulence role in the fate of virus-containing droplets in violent expiratory events. Physical Review Research, 3(1):013091, 2021.

\bibitem{Ro21} M. E. Rosti, S. Olivieri, M. Cavaiola, A. Seminara, and A. Mazzino. Fluid dynamics of COVID-19 airborne infection suggests urgent data for a scientific design of social distancing. Scientific Reports, 10(1):1–9, 2020.

\bibitem{Lo21} K. L. Chong, C. S. Ng, N. Hori, R. Yang, R. Verzicco, and D. Lohse. Extended lifetime of respiratory droplets in a turbulent vapor puff and its implications on airborne disease transmission. Physical Review Letters, 126(3):034502, 2021.

\bibitem{Ko41}  A.N. Kolmogorov, The local structure of turbulence in incompressible viscous fluid for very large Reynolds numbers. C.R. Acad. Sci. URSS 30, 301-305 (1941).

\bibitem{Ob41} A. Obukhov, On the distribution of energy in the spectrum of turbulent flow. C.R. Acad. Sci. URSS32, 22-24 (1941).

\bibitem{Bi03}  L. Biferale, G. Boffetta, A. Celani, A. Lanotte, F. Toschi, and M. Vergassola, The decay of homogeneous anisotropic turbulence. Phys Fluids 15, 2105 (2003).

\bibitem{Tr88} D. Tritton, Physical Fluid Dynamics. Oxford, UK: Clarendon 
(1988).

\bibitem{Ko75}  L.S. Kovasznay, H. Fujita, R.L. Lee, Unsteady turbulent puffs, in Advances in Geophysics 18, 253-263 (Elsevier, Amsterdam, 1975).

\bibitem{PRL}  A. Mazzino, M.  E. Rosti, Unraveling the secrets of turbulence in a fluid puff. Phys. Rev. Lett. in press (2021).

\bibitem{Gh10}  A. Ghaem-Maghami, H. Johari, Velocity field of isolated turbulent puffs. Phys Fluids 22, 115105 (2010).

\bibitem{Ch03} M. Chertkov, Phenomenology of Rayleigh-Taylor turbulence. Phys. Rev. Lett. 91, 115001 (2003).

\bibitem{Vl09} N. Vladimirova, M. Chertkov, Self-similarity and universality in Rayleigh-Taylor, Boussinesq turbulence. Phys. Fluids 21, 015102 (2009).

\bibitem{Bo09} G. Boffetta, A. Mazzino, S. Musacchio, L. Vozella, Kolmogorov scaling and intermittency in Rayleigh-Taylor turbulence, Phys. Rev. E 
79, 065301(R) (2009).

\bibitem{Ce06}  A. Celani, A. Mazzino, L. Vozella, Rayleigh-Taylor Turbulence in Two Dimensions, Phys. Rev. Lett. 96, 134504 (2006).
  
\bibitem{Bo17}  G. Boffetta, A. Mazzino, Incompressible Rayleigh-Taylor turbulence, Ann. Rev. Fluid Mech. 49, 119-143 (2017).

\bibitem{Ob49}  A. Obukhov, Temperature field structure in a turbulent flow. Izv. Acad. Nauk SSSR Geogr. Geophys. 13, 58-69 (1949).

\bibitem{Co51} S. Corrsin, On the spectrum of isotropic temperature fluctuations in an isotropic turbulence. J. Appl. Phys. 22, 469-473 (1951).

\bibitem{Fr95}  U. Frisch, Turbulence - The legacy of AN Kolmogorov, (1995).

\bibitem{Fr98}  U. Frisch, A. Mazzino, M. Vergassola, Intermittency in passive scalar advection, Phys. Rev. Lett. 80, 5532-5537 (1998).

\bibitem{Ve97} M. Vergassola, A. Mazzino, Structures and intermittency in 
a passive scalar model, Phys. Rev. Lett. 79, 1849-1852 (1997).

\bibitem{Sh00} B.I. Shraiman, E.D. Siggia, Scalar turbulence, Nature 405, 
639-646 (2000).

\bibitem{Fa01} G. Falkovich, K. Gawkedzki, M. Vergassola, Particles and fields in fluid turbulence, Rev. Mod. Phys. 73, 913-975 (2001).

\bibitem{Wa07} T. Watanabe,  T. Gotoh, Inertial-range intermittency and accuracy of direct numerical simulation for turbulence and passive scalar turbulence. J. Fluid Mech. 590, 117-146 (2007).

\bibitem{rosti_brandt_2017a} M.E. Rosti, L. Brandt, Numerical simulation of turbulent channel flow over a viscous hyper-elastic wall. Journal of Fluid Mechanics 830, 708–735 (2017).

\bibitem{rosti2019flowing} M.E. Rosti, S. Olivieri, A.A. Banaei, L. Brandt, A. Mazzino, Flowing fibers as a proxy of turbulence statistics. Meccanica 55, 357–370 (2020).

\bibitem{Al19} D. Alghalibi, M. E. Rosti, and L. Brandt. Inertial migration of a deformable particle in pipe flow. Physical Review Fluids, 4(10):104201 (2019).

\bibitem{rosti_ge_jain_dodd_brandt_2019} M.E. Rosti, Z. Ge, S.S. Jain, M.S. Dodd, L. Brandt, Droplets in homogeneous shear turbulence. Journal of Fluid Mechanics 876, 962–984 (2019).

\bibitem{rosti2020increase} M.E. Rosti, L. Brandt, Increase of turbulent drag by polymers in particle suspensions. Physical Review Fluids 5, 041301(R) (2020).

\bibitem{olivieri2020dispersed} S. Olivieri, L. Brandt, M.E. Rosti, A. Mazzino, Dispersed fibers change the classical energy budget of turbulence via nonlocal transfer. Physical Review Letters 125, 114501 (2020).

\bibitem{devita2020} F. De Vita, M. E. Rosti, S. Caserta, and L. Brandt. Numerical simulations of vorticity banding of emulsions in shear flows. Soft Matter, 16:2854–2863, 2020.

\bibitem{devita2019} F. De Vita, M. E. Rosti, S. Caserta, and L. Brandt. On the effect of coalescence on the rheology of emulsions. Journal of Fluid Mechanics, 880:969–991, 2019.

\bibitem{Be93} R. Benzi, S. Ciliberto, R. Tripiccione, C. Baudet, S. Massaioli, S. Succi, Extended self-similarity in turbulent flows. Phys. Rev. E 48, R29(R) (1993).

\bibitem{Ch10} S. Chakraborty, U. Frisch, and S. S. Ray. Extended self-similarity works for the Burgers equation and why. Journal of Fluid Mechanics, 649:275–285, 2010.

\bibitem{Ca95}  V. Carbone, P. Veltri,  and R. Bruno,  Experimental evidence for differences in the extended self-similarity scaling laws between fluid and magnetohydrodynamic turbulent flows.  Phys. Rev. Lett. 75, 3110 (1995).

\bibitem{Be96} R. Benzi, L. Biferale,  S. Ciliberto, M. V.  Struglia,  R. Tripiccione,  Scaling property of turbulent flows, Phys.  Rev.  E, 53, R3025~1996

\bibitem{Me96} C.  Meneveau,  Transition between viscous and inertial-range scaling of turbulence structure functions, Phys.  Rev.  E, 54, 3657~1996

\bibitem{Le99}  G. S.  Lewis and H.  L.  Swinney, Phys.  Rev.  E, 59, 5457 (1999)

\bibitem{Fr78} U. Frisch, P. L. Sulem, and M. Nelkin. A simple dynamical model of intermittent fully developed turbulence. Journal of Fluid Mechanics, 87(4):719–736, 1978.

\bibitem{Ma06} A. Celani, A. Mazzino, and L. Vozella. Rayleigh-Taylor turbulence in two dimensions. Physical review letters 96.13 (2006): 134504.

\end{thebibliography}
\end{document}